\documentstyle[12pt]{article}
\date{}
\author{Valerii Dryuma,\\[5mm]
{\it Institute of Mathematics and Informatics, AS RM,}\\[3mm] {\it
5 Academiei Street, 2028 Kishinev, Moldova},
 }
\title{ON SOLUTIONS OF A HEAVENLY EQUATIONS \\[2mm]AND THEIR GENERALIZATIONS}

\begin{document}
\maketitle

\begin{abstract}
\ \ \ \ The solutions of the Heavenly Equations and their
generalizations are considered.
\end{abstract}

\section{ Method of solution}

      To integrate the partial nonlinear first order differential equation
\begin{equation}\label{dryuma:eq1}
F(x,y,z,w,f_x,f_y,f_z,f_w,f_{xx},f_{xy},f_{xz},f_{yy},..)=0
 \end{equation}
    can be applied a following approach \cite{dryuma:dr1}.

   We use the parametric presentation of the functions and variables
\begin{equation}\label{dryuma:eq2}
f(x,y,z,w)\rightarrow u(x,t,z,w),\quad y \rightarrow
v(x,t,z,w),\quad f_x\rightarrow u_x-\frac{u_t}{v_t}v_x,\]\[\quad
f_w\rightarrow u_w-\frac{u_t}{v_t}v_w,\quad f_z\rightarrow
u_z-\frac{u_t}{v_t}v_z,\quad f_y \rightarrow \frac{u_t}{v_t},
\end{equation}
where variable $t$ is considered as parameter.

  Remark that conditions of the type
   \[
   f_{xt}=f_{tx},\quad f_{xz}=f_{zx},\quad f_{xw}=f_{wx}...
   \]
are fulfilled at such type of presentation.

  In result instead of equation (\ref{dryuma:eq1}) one get the
  relation between the new variables $u(x,t,z,w)$ and $v(x,t,z,w)$ and
  their partial derivatives
\begin{equation}\label{dryuma:eq3}
\Phi(u,v,u_x,u_z,u_w,u_t,v_x,v_z,v_w,v_t...)=0.
  \end{equation}

    In some cases the solution of such type of indefinite equation is more simple
    problem than the solution of the equation (\ref{dryuma:eq1}).

    We apply this method to integrating of the  equations having
    applications in various branches of modern mathematical physics.

\section{The first heavenly equation}

     The Plebanski first heavenly equation
\begin{equation}\label{dryuma:eq4}
 \left ({\frac {\partial ^{2}}{\partial x\partial y}}\theta(x,y,z,w)
\right ){\frac {\partial ^{2}}{\partial w\partial
z}}\theta(x,y,z,w)- \left ({\frac {\partial ^{2}}{\partial
w\partial x}}\theta(x,y,z,w) \right ){\frac {\partial
^{2}}{\partial y\partial z}}\theta(x,y,z,w)-1 =0
\end{equation}
arise in connection with the 4-dimensional spaces endowed with the
metrics
\begin{equation} \label{dryuma:eq5}
ds^2=2\theta_{xy}dxdy+2\theta_{zw}dzdw+2\theta_{xw}dxdw+2\theta_{zy}dzdy
\end{equation}
which are simultaneously Ricci flat and self-dual (\cite{Boy2})

  In result of applying of the transformations
\begin{equation}\label{dryuma:eq6}
\theta(x,y,z,w)=u(x,t,z,w),\quad y=v(x,t,z,w)
\end{equation}
 and
\begin{equation}\label{dryuma:eq7}
\frac{\partial\theta(x,y,z,w)}{\partial x}={\frac {\partial
}{\partial x}}u(x,t,z,w)-{\frac {\left ({\frac {
\partial }{\partial t}}u(x,t,z,w)\right ){\frac {\partial }{\partial x
}}v(x,t,z,w)}{{\frac {\partial }{\partial t}}v(x,t,z,w)}},
\end{equation}
\[
\frac{\partial\theta(x,y,z,w)}{\partial z}={\frac {\partial
}{\partial z}}u(x,t,z,w)-{\frac {\left ({\frac {
\partial }{\partial t}}u(x,t,z,w)\right ){\frac {\partial }{\partial z
}}v(x,t,z,w)}{{\frac {\partial }{\partial t}}v(x,t,z,w)}},
\]
\[
\frac{\partial\theta(x,y,z,w)}{\partial w}={\frac {\partial
}{\partial w}}u(x,t,z,w)-{\frac {\left ({\frac {
\partial }{\partial t}}u(x,t,z,w)\right ){\frac {\partial }{\partial w
}}v(x,t,z,w)}{{\frac {\partial }{\partial t}}v(x,t,z,w)}},
\]
\[
\frac{\partial\theta(x,y,z,w)}{\partial t}={\frac {{\frac
{\partial }{\partial t}}u(x,t,z,w)}{{\frac {\partial }{
\partial t}}v(x,t,z,w)}}
\]
with corresponding expressions for the second order derivatives at
the equation (\ref{dryuma:eq4}) one gets the relation between the
functions $u(x,t,z,w)$, $v(x,t,z,w)$ and their partial derivatives
\begin{equation}\label{dryuma:eq8}
F(u,v,u_x,u_t,u_z,u_w,...)=0. \end{equation}

The  corresponding explicit expression for the function $F$ can be
obtained with the help of the MAPLE and we omit its  in view of
its inconvenience.

    To cite as an example some of the simplest reduction of full
    expression ~(\ref{dryuma:eq8}).

    The substitution
    \[
    v(x,t,z,w)=t{\frac {\partial }{\partial t}}\omega(x,t,z,w)-\omega(x,t,
z,w),
\]
\[
u(x,t,z,w)={\frac {\partial }{\partial t}}\omega(x,t,z,w)
\]
give us the equation
\begin{equation}\label{dryuma:eq9}\left ({\frac {\partial ^{2}}{\partial w\partial x}}\omega(x,t,z,w)
\right ){\frac {\partial }{\partial
z}}\omega(x,t,z,w)-{t}^{4}{\frac {
\partial ^{2}}{\partial {t}^{2}}}\omega(x,t,z,w)+\]\[+t\left ({\frac {
\partial ^{2}}{\partial t\partial x}}\omega(x,t,z,w)\right ){\frac {
\partial ^{2}}{\partial w\partial z}}\omega(x,t,z,w)-\left ({\frac {
\partial ^{2}}{\partial w\partial x}}\omega(x,t,z,w)\right )t{\frac {
\partial ^{2}}{\partial t\partial z}}\omega(x,t,z,w)-\]\[-\left ({\frac {
\partial }{\partial x}}\omega(x,t,z,w)\right ){\frac {\partial ^{2}}{
\partial w\partial z}}\omega(x,t,z,w)=0.
\end{equation}

It has the particular  solution
\[
\omega(x,t,z,w)=z+{e^{x+w}}t\left ({\it
\_C1}\,\sinh({t}^{-1})+{\it \_C2}\,\cosh({t}^{-1})\right )
\]
where $\_C1$ and $\_C2$ are the parameters.

    After substitution of this expression into the relations
    \[
    y-t{\frac {\partial }{\partial
    t}}\omega(x,t,z,w)+\omega(x,t,z,w)=0, \quad \theta(x,y,z,w)-{\frac {\partial }{\partial
    t}}\omega(x,t,z,w)=0
\]
we find the parametric presentation of variables (in particular
case $\_C2=0$)
\[
y+{e^{x+w}}{\it \_C1}\,\sinh({t}^{-1})+z=0
\]
and
\[
\theta(x,y,z,w)t-{e^{x+w}}{\it
\_C1}\,t\cosh({t}^{-1})+{e^{x+w}}{\it \_C1}\,\sinh({t}^{-1})=0.
\]

Elimination of the parameter $t$ from both relations give us the
explicit expression for the function
\[
\theta(x,y,z,w)={\it \_C1}\,\sqrt {{\frac
{{e^{-2\,x-2\,w}}{y}^{2}+2\,
{e^{-2\,x-2\,w}}yz+{e^{-2\,x-2\,w}}{z}^{2}+{{\it \_C1}}^{2}}{{{\it
\_C1}}^{2}}}}{e^{x+w}}-\]\[(y+z)~{\it arcsinh}({\frac {\left
(y+z\right ){e^{-x-w} }}{{\it \_C1}}})
\]
which is solution of the first heavenly equation
(\ref{dryuma:eq4}).

    The substitution
    \[
    u(x,t,z,w)=t{\frac {\partial }{\partial t}}\omega(x,t,z,w)-\omega(x,t,
z,w),
\]
\[
v(x,t,z,w)={\frac {\partial }{\partial t}}\omega(x,t,z,w)
\]
into the expression  ~(\ref{dryuma:eq8}) lead to the equation
\begin{equation}\label{dryuma:eq10}
\left ({\frac {\partial ^{2}}{\partial t\partial
x}}\omega(x,t,z,w) \right ){\frac {\partial ^{2}}{\partial
w\partial z}}\omega(x,t,z,w)- \left ({\frac {\partial
^{2}}{\partial t\partial z}}\omega(x,t,z,w) \right ){\frac
{\partial ^{2}}{\partial w\partial x}}\omega(x,t,z,w)-\]\[-{ \frac
{\partial ^{2}}{\partial {t}^{2}}}\omega(x,t,z,w) =0.
\end{equation}

  It has solution in form
  \[
  \omega(x,t,z,w)=A(x,t,z)w
\]
where
\[
\left ({\frac {\partial ^{2}}{\partial t\partial x}}A(x,t,z)\right
){ \frac {\partial }{\partial z}}A(x,t,z)-\left ({\frac {\partial
^{2}}{
\partial t\partial z}}A(x,t,z)\right ){\frac {\partial }{\partial x}}A
(x,t,z)-\]\[-{\frac {\partial ^{2}}{\partial {t}^{2}}}A(x,t,z)=0.
\]

    From here we find a simplest solution
    \[
    A(x,t,z)=\ln ({\frac {t}{z}}-1)+x.
\]

      In result we get the relations
      \[
t={\frac {yz+w}{y}}
\]
and
\[
\theta(x,y,z,w)=-w\left (-t+\ln ({\frac {t-z}{z}})t-\ln ({\frac
{t-z}{ z}})z+xt-xz\right )\left (t-z\right )^{-1}.
\]

Elimination of parameter $t$ from these relations give us  the
explicit expression for the function
\[
\theta(x,y,z,w)=yz+w-w\ln ({\frac {w}{yz}})-wx
\]
which is solution of the first heavenly equation
(\ref{dryuma:eq4}).

    By analogy can be constructed a more complicated solutions of the equation (\ref{dryuma:eq4}).

\section{The second heavenly equation}

     The Plebanski second heavenly equation has the form
\begin{equation}\label{dryuma:eq11}
{\frac {\partial ^{2}}{\partial w\partial
x}}\theta(x,y,z,w)+{\frac {
\partial ^{2}}{\partial y\partial z}}\theta(x,y,z,w)+\left ({\frac {
\partial ^{2}}{\partial {x}^{2}}}\theta(x,y,z,w)\right ){\frac {
\partial ^{2}}{\partial {y}^{2}}}\theta(x,y,z,w)-\]\[-\left ({\frac {
\partial ^{2}}{\partial x\partial y}}\theta(x,y,z,w)\right
)^{2}=0.
\end{equation}

   It describes the properties of the  Ricci flat a 4-dimensional
   spaces endowed with the metrics
\begin{equation}\label{dryuma:eq12}
ds^2=-\theta_{xx}dz^2+2\theta_{xy}dw dz-\theta_{yy}dw^2+dw dx+dz
dy \end{equation}
 satisfying  the
    conditions of self-duality for corresponding Riemann tensor
    $R_{i j k l}$ of the space \cite{Dun1}.

    To integrate the equation (\ref{dryuma:eq11}) we transform its
    in another form with accordance of the rules
    (\ref{dryuma:eq2}).

    In result of such transformation one gets the relation between
    the functions $u(x,t,z,w)$ and $v(x,t,z,w)$ and their
    partial derivatives
\begin{equation}\label{dryuma:eq13}
F(u,v,u_x,u_t,u_z,u_w,u_{xx},u_{xt},u_{xz}....)=0.
\end{equation}

    The next step is receiving  the p.d.e from this relation using  the substitutions
    similar to
    \[
    u(x,t,z,w)=t~\frac{\partial}{\partial t}\omega(x,t,z,w)-\omega(x,t,z,w),\quad
    v(x,t,z,w)=\frac{\partial}{\partial t}\omega(x,t,z,w),
    \]
or
    \[
    v(x,t,z,w)=t~\frac{\partial}{\partial t}\omega(x,t,z,w)-\omega(x,t,z,w),\quad
    u(x,t,z,w)=\frac{\partial}{\partial t}\omega(x,t,z,w),
    \]
or the more complicated.

     As example we get from the relation (\ref{dryuma:eq13}) the equation
\begin{equation}\label{dryuma:eq14}-\left ({\frac {\partial ^{2}}{\partial {t}^{2}}}\omega(x,t,z,w)
\right ){\frac {\partial ^{2}}{\partial w\partial
x}}\omega(x,t,z,w)-{ \frac {\partial ^{2}}{\partial
{x}^{2}}}\omega(x,t,z,w)-{\frac {
\partial ^{2}}{\partial t\partial z}}\omega(x,t,z,w)+\]\[+\left ({\frac {
\partial ^{2}}{\partial t\partial w}}\omega(x,t,z,w)\right ){\frac {
\partial ^{2}}{\partial t\partial x}}\omega(x,t,z,w)=0
\end{equation}
 with the help of the first substitution.

   The simplest solution of this equation is
\[
\omega(x,t,z,w)=-{\frac {{t}^{2}w}{x}}+z
\]
and corresponds the solution of the equation (\ref{dryuma:eq11})
\[
\theta(x,y,z,w)=-1/4\,{\frac {{y}^{2}x+4\,zw}{w}}
\]
after the return  at the assumed function $\theta(x,y,z,w)$.

   The substitution
   \[
   \omega(x,t,z,w)=B(t,z,w)+xC(t,z,w)
\]
into the equation (\ref{dryuma:eq14}) give us the system
\[
-\left ({\frac {\partial }{\partial w}}C(t,z,w)\right ){\frac {
\partial ^{2}}{\partial {t}^{2}}}C(t,z,w)+\left ({\frac {\partial }{
\partial t}}C(t,z,w)\right ){\frac {\partial ^{2}}{\partial t\partial
w}}C(t,z,w)-{\frac {\partial ^{2}}{\partial t\partial
z}}C(t,z,w)=0
\]
\[-\left ({\frac {\partial }{\partial w}}C(t,z,w)\right ){\frac {
\partial ^{2}}{\partial {t}^{2}}}B(t,z,w)+\left ({\frac {\partial }{
\partial t}}C(t,z,w)\right ){\frac {\partial ^{2}}{\partial t\partial
w}}B(t,z,w)-{\frac {\partial ^{2}}{\partial t\partial
z}}B(t,z,w)=0.
\]

   Its simplest solution is
   \[C(t,z,w)={\frac {\left ({e^{t}}+1\right )w}{z}}
\]
and
\[B(t,z,w)=w\left ({\it \_F2}(z)+\int \!{\it \_F1}(z\left (1+{e^
{-t}}\right )){e^{t}}{dt}+\int \!{\it \_F1}(z\left
(1+{e^{-t}}\right ) ){dt}\right ) ,
\]
where $\_Fi(z)$ are  arbitrary.

    Using these expressions we get the function $\omega(x,t,z,w)$ from
    (\ref{dryuma:eq14}).

    In particular case we have
        \[
    \omega(x,t,z,w)=\left ({\it \_F2}(z)+z{e^{t}}+2\,zt-z{e^{-t}}\right )w
+{\frac {x\left ({e^{t}}+1\right )w}{z}}.
\]

   Using this expression and the relations
\[
y-{\frac {\partial }{\partial t}}\omega(x,t,z,w) =0,
\]
\[
\theta(x,y,z,w)-t{\frac {\partial }{\partial
t}}\omega(x,t,z,w)+\omega (x,t,z,w)=0
\]
we find
\[
yz-w{z}^{2}{e^{t}}-2\,w{z}^{2}-w{z}^{2}{e^{-t}}-wx{e^{t}} =0
\]
and
\[\left (-tw{e^{t}}-tw{e^{-t}}+w{e^{t}}-w{e^{-t}}\right )z+\theta(x,y,z,
w)+w{\it \_F2}(z)+{\frac {wx-twx{e^{t}}+wx{e^{t}}}{z}} =0.
\]

    Elimination of the parameter $t$ from these relations give us
    the function  $\theta(x,y,z,w)$
which is solution of the second heavenly equation
 (\ref{dryuma:eq11}).

\section{The Dunajski heavenly equation}

     The Dunajski heavenly equation has the form of the system
\begin{equation}\label{dryuma:eq15}
{\frac {\partial ^{2}}{\partial w\partial x}}f(x,y,z,w)+{ \frac
{\partial ^{2}}{\partial y\partial z}}f(x,y,z,w)+\]\[+\left
({\frac {
\partial ^{2}}{\partial {x}^{2}}}f(x,y,z,w)\right ){\frac {\partial ^{
2}}{\partial {y}^{2}}}f(x,y,z,w)-\left ({\frac {\partial ^{2}}{
\partial x\partial y}}f(x,y,z,w)\right )^{2}=L(x,y,z,w),
\]
\\[1mm]\[
{\frac {\partial ^{2}}{\partial w\partial x}}L(x,y,z,w)+{\frac {
\partial ^{2}}{\partial y\partial z}}L(x,y,z,w)+\left ({\frac {
\partial ^{2}}{\partial {x}^{2}}}f(x,y,z,w)\right ){\frac {\partial ^{
2}}{\partial {y}^{2}}}L(x,y,z,w)+\]\[+\left ({\frac {\partial
^{2}}{
\partial {y}^{2}}}f(x,y,z,w)\right ){\frac {\partial ^{2}}{\partial {x
}^{2}}}L(x,y,z,w)-2\,\left ({\frac {\partial ^{2}}{\partial
x\partial y}}f(x,y,z,w)\right ){\frac {\partial ^{2}}{\partial
x\partial y}}L(x, y,z,w)=0.
\end{equation}

It describes  the properties of a four-dimensional Riemannian
spaces having the metrics
\begin{equation}\label{dryuma:eq16}
ds^2=-f_{xx}dz^2+2f_{xy}dw dz-f_{yy}dw^2+dw dx+dz dy
\end{equation}
 with condition on the Ricci-tensor  of the metric (\ref{dryuma:eq16})
 $$
 R_{ij}=T_{ij},
 $$
  where  $T_{ij}$ is a Maxwell stress-energy tensor of electromagnetic field \cite{Dun2}.

 The conditions of
self-duality for the Riemann tensor $R_{i j k l}$ of these spaces
are equivalent the equation (\ref{dryuma:eq15}).

   In connection with this  it is interested to note that the equation (\ref{dryuma:eq15})
 coincides with condition
 \[
 C_{1212}=0,
 \]
for the component $C_{1212}$ of the Weyl tensor of the metrics
(\ref{dryuma:eq16}).

     To construction of the solutions of the equation (\ref{dryuma:eq15}) we
     transform its in accordance with the rules (\ref{dryuma:eq2}).

     In result we get the relation
\begin{equation}\label{dryuma:eq17}
\Phi(u,v,u_x,u_t,u_z,u_w,u_{xx},u_{xt},u_{xz}....)=0
\end{equation}
which is reduced to the  p.d.e.'s when between the $u$ and $v$
exist functional dependence.

     Let us consider an examples.

     In the case
\[
u(x,t,z,w)=t{\frac {\partial }{\partial
t}}\omega(x,t,z,w)-\omega(x,t, z,w),\quad v(x,t,z,w)={\frac
{\partial }{\partial t}}\omega(x,t,z,w)
\]
one get the p.d.e. on the function $\omega(x,t,z,w)$
\[
M(\omega,\omega_x,\omega_t,\omega_z,\omega_w,...)=0.
\]

    Its simplest solution can be presented in form
\[
\omega(x,t,z,w)=A(t)+xzt+w,
\]
where the function $A(t)$ is defined by the equation
 \[
-3\,\left ({\frac {d^{3}}{d{t}^{3}}}A(t)\right )^{2}+\left ({\frac
{d^ {2}}{d{t}^{2}}}A(t)\right ){\frac {d^{4}}{d{t}^{4}}}A(t)=0.
\]

    In result we get the solution
    \[
    \omega(x,t,z,w)=2/3\,{\frac {\sqrt {2}\left (-t{\it \_C1}-{\it \_C2}\,
{\it \_C1}\right )^{3/2}}{{{\it \_C1}}^{2}}}+{\it \_C3}\,t+{\it
\_C4}+ xzt
\]
from which is followed
\[
t=\]\[-{\it \_C2}-1/2\,{\it \_C1}\,{y}^{2}+{\it \_C1}\,y{\it
\_C3}+{\it \_C1 }\,yxz-1/2\,{\it \_C1}\,{{\it \_C3}}^{2}-{\it
\_C1}\,{\it \_C3}\,xz-1/ 2\,{\it \_C1}\,{x}^{2}{z}^{2} ,
\]
and
\[f(x,y,z,w)=\]\[-1/3\,{\frac {-{t}^{2}\sqrt {2}+t\sqrt {2}{\it \_C2}+2\,
\sqrt {2}{{\it \_C2}}^{2}+3\,{\it \_C4}\,\sqrt {-{\it \_C1}\,\left
(t+ {\it \_C2}\right )}+3\,w\sqrt {-{\it \_C1}\,\left (t+{\it
\_C2}\right )}}{\sqrt {-{\it \_C1}\,\left (t+{\it \_C2}\right )}}}
.
\]

Elimination of the parameter $t$ from these relations give us the
function
\[
 f(x,y,z,w)=1/6\,{\it \_C1}\,{x}^{3}{z}^{3}+\left
(-1/2\,{\it \_C1}\,{x }^{2}y+1/2\,{\it \_C3}\,{\it
\_C1}\,{x}^{2}\right ){z}^{2}+\]\[+\left ( \left (1/2\,{\it
\_C1}\,{{\it \_C3}}^{2}+{\it \_C2}\right )x+1/2\,x{ \it
\_C1}\,{y}^{2}-{\it \_C1}\,{\it \_C3}\,xy\right )z-1/6\,{y}^{3}{
\it \_C1}+1/2\,{\it \_C3}\,{\it \_C1}\,{y}^{2}-\]\[-w+\left
(-1/2\,{\it \_C1}\,{{\it \_C3}}^{2}-{\it \_C2}\right )y+1/6\,{\it
\_C1}\,{{\it \_C3}}^{3}+{\it \_C3}\,{\it \_C2}-{\it \_C4}
\]
 satisfying  the equation (\ref{dryuma:eq15}).

The substitution
    \[
v(x,t,z,w)=t{\frac {\partial }{\partial
t}}\omega(x,t,z,w)-\omega(x,t, z,w),\quad u(x,t,z,w)={\frac
{\partial }{\partial t}}\omega(x,t,z,w)
\]
lead to the p.d.e. on the function $\omega(x,t,z,w)$
\[
N(\omega,\omega_x,\omega_t,\omega_z,\omega_w,...)=0.
\]
 having the solution
 \[
\omega(x,t,z,w)=A{e^{t}}w+\left (x+z\right )t.
\]

    This solution give us the following presentation
    \[y-tA{e^{t}}w+A{e^{t}}w=0
\]
and
\[f(x,y,z,w)-A{e^{t}}w-x-z=0.
\]

Elimination of the parameter $t$ from these relations give us the
function
\[
f(x,y,z,w)={e^{{\it LambertW}({\frac {y{e^{-1}}}{Aw}})+1}}Aw+x+z
\]
or \[f(x,y,z,w)=\]\[= \left(y+x{\it LambertW}({\frac
{y{e^{-1}}}{Aw}})+z{\it LambertW}({ \frac
{y{e^{-1}}}{Aw}})\right)\left ({\it LambertW}({\frac {y{e^{-1}}}
{Aw}})\right )^{-1},
\]
satisfying  the equation (\ref{dryuma:eq15}).

The LambertW function in these formulas is defined by the equation
\[
      LambertW(x)  exp(LambertW(x)) = x .
\]

 From here it is apparent how to construct another examples of solutions of
  the equation (\ref{dryuma:eq15}).
\section{Six-dimensional generalization of the first heavenly
equation}

    The first heavenly equation in dimension Dim=6 has the form
\begin{equation}\label{dryuma:eq18}
\left ({\frac {\partial ^{2}}{\partial u\partial z}}A(\vec x )
\right )\left ({\frac {\partial ^{2}}{\partial v\partial w}}A(\vec
x)\right ){\frac {\partial ^{2}}{\partial x\partial y}}A(\vec x
)-\left ({\frac {\partial ^{2}}{\partial u\partial z}}A(\vec x)
\right )\left ({\frac {\partial ^{2}}{\partial w\partial x}}A(\vec
x)\right ){\frac {\partial ^{2}}{\partial v\partial y}}A(\vec x
)+\]\[+\left ({\frac {\partial ^{2}}{\partial y\partial z}}A(\vec
x) \right )\left ({\frac {\partial ^{2}}{\partial w\partial
x}}A(\vec x)\right ){\frac {\partial ^{2}}{\partial u\partial
v}}A(\vec x )-\left ({\frac {\partial ^{2}}{\partial w\partial
z}}A(\vec x) \right )\left ({\frac {\partial ^{2}}{\partial
u\partial v}}A(\vec x)\right ){\frac {\partial ^{2}}{\partial
x\partial y}}A(\vec x )-\]\[-\left ({\frac {\partial
^{2}}{\partial y\partial z}}A(\vec x) \right )\left ({\frac
{\partial ^{2}}{\partial u\partial x}}A(\vec x)\right ){\frac
{\partial ^{2}}{\partial v\partial w}}A(\vec x)+\left ({\frac
{\partial ^{2}}{\partial u\partial x}}A(\vec x) \right )\left
({\frac {\partial ^{2}}{\partial v\partial y}}A(\vec x)\right
){\frac {\partial ^{2}}{\partial w\partial z}}A(\vec x)-\]\[-1=0,
\end{equation}
where $A(\vec x)=A(x,y,z,u,v,w)$.

     It arises as condition on the six-dimensional metrics
         \[
     {{\it ds}}^{2}=2\,\left ({\frac {\partial ^{2}}{\partial x\partial y}}
A(\vec x)\right ){\it dx}\,{\it dy}+2\,\left ({\frac {\partial ^{
2}}{\partial u\partial x}}A(\vec x)\right ){\it dx}\,{\it du}+2\,
\left ({\frac {\partial ^{2}}{\partial w\partial x}}A(\vec x)
\right ){\it dx}\,{\it dw}+\]\[+2\,\left ({\frac {\partial
^{2}}{\partial y
\partial z}}A(\vec x)\right ){\it dy}\,{\it dz}+2\,\left ({\frac
{\partial ^{2}}{\partial v\partial y}}A(\vec x)\right ){\it dy}\,
{\it dv}+2\,\left ({\frac {\partial ^{2}}{\partial u\partial
z}}A(\vec x)\right ){\it dz}\,{\it du}+\]\[+2\,\left ({\frac
{\partial ^{2}}{
\partial w\partial z}}A(\vec x)\right ){\it dz}\,{\it dw}+2\,
\left ({\frac {\partial ^{2}}{\partial u\partial v}}A(\vec x)
\right ){\it du}\,{\it dv}+2\,\left ({\frac {\partial
^{2}}{\partial v
\partial w}}A(\vec x)\right ){\it dv}\,{\it dw}.
\]
 to be a Ricci-flat
\[
R_{ij}=0.
\]

     The transformation of the equation (\ref{dryuma:eq18}) at the new form
     in accordance with the rules (\ref{dryuma:eq2}) lead to the relation
     \begin{equation}\label{dryuma:eq19}
\Psi(U,V,U_x,U_t,U_z,U_w,U_w,U_{xx},U_{xt},U_{xz}...,V_x,V_t,V_z,V_w,V_w,V_{xx},V_{xt},V_{xz}...)=0.
\end{equation}

    This relation moves to the p.d.e.
\begin{equation}\label{dryuma:eq20}
-\left ({\frac {\partial ^{2}}{\partial t\partial z}}\omega(\vec
\tau)\right )\left ({\frac {\partial ^{2}}{\partial w\partial
x}}\omega(\vec \tau)\right ){\frac {\partial ^{2}}{\partial
u\partial v}}\omega (\vec \tau)-\left ({\frac {\partial
^{2}}{\partial t\partial x}} \omega(\vec \tau)\right )\left
({\frac {\partial ^{2}}{\partial u
\partial z}}\omega(\vec
\tau)\right ){\frac {\partial ^{2}}{
\partial v\partial w}}\omega(\vec
\tau)+\]\[+\left ({\frac {\partial ^{2} }{\partial t\partial
z}}\omega(\vec \tau)\right )\left ({\frac {
\partial ^{2}}{\partial u\partial x}}\omega(\vec
\tau)\right ){ \frac {\partial ^{2}}{\partial v\partial
w}}\omega(\vec \tau)+\left ({\frac {\partial ^{2}}{\partial
t\partial v}}\omega(\vec \tau) \right )\left ({\frac {\partial
^{2}}{\partial u\partial z}}\omega(\vec \tau)\right ){\frac
{\partial ^{2}}{\partial w\partial x}}\omega(\vec \tau)+\]\[+\left
({\frac {\partial ^{2}}{\partial t\partial x}}\omega (\vec
\tau)\right )\left ({\frac {\partial ^{2}}{\partial w\partial
z}}\omega(\vec \tau)\right ){\frac {\partial ^{2}}{\partial u
\partial v}}\omega(\vec
\tau)-\left ({\frac {\partial ^{2}}{\partial t
\partial v}}\omega(\vec
\tau)\right )\left ({\frac {\partial ^{2}}{
\partial u\partial x}}\omega(\vec
\tau)\right ){\frac {\partial ^{2} }{\partial w\partial
z}}\omega(\vec \tau)-\]\[-{\frac {\partial ^{2}}{\partial {t}^{2
}}}\omega(\vec \tau)=0 \end{equation}  at the substitutions
\[U(x,t,z,u,v,w)=t{\frac {\partial }{\partial
t}}\omega(x,t,z,u,v,w)- \omega(x,t,z,u,v,w),\]\[
V(x,t,z,u,v,w)={\frac {\partial }{\partial t}}\omega(x,t,z,u,v,w),
\]
where $\omega(\vec \tau)= \omega(x,t,z,u,v,w)$.

    The solutions of last equation allow us to construct the solutions of initial equation
    (\ref{dryuma:eq18}).

    Let us consider an examples.

    The solution of the equation (\ref{dryuma:eq20}) of the form
    \[
    \omega(x,t,z,u,v,w)=Bxt+uv+zw+1/2\,B{t}^{2}
\]
corresponds the solution of equation (\ref{dryuma:eq18})
\[
A(x,y,z,t,u,v,w)=-1/2\,{\frac
{-{y}^{2}+2\,yBx-{B}^{2}{x}^{2}+2\,uvB+2 \,zwB}{B}}.
\]

    The solution of the form
\[\omega(x,t,z,u,v,w)=\ln ({\frac {{\it \_C1}\,\left (xz+t\right )}{z}}+
{\it \_C2})u+vw
\]
lead to the solution of equation (\ref{dryuma:eq18})
\[
A(x,y,z,t,u,v,w)=-\left(y{\it \_C1}\,xz+y{\it \_C2}\,z-{\it
\_C1}\,u+ \ln ({\frac {{\it \_C1}\,u}{yz}})u{\it \_C1}+vw{\it
\_C1}\right){{\it \_C1}}^{-1}.
\]

    The solution of the equation (\ref{dryuma:eq20}) of the form
    \[\omega(x,t,z,u,v,w)=\sinh(x+t)\left (zu+vw\right )
\]
give us the solution of equation (\ref{dryuma:eq18})
\[
A(x,y,z,t,u,v,w)=-\sqrt {-{\frac {-y+zu+vw}{zu+vw}}}\sqrt {{\frac
{y+z u+vw}{zu+vw}}}vw-yx+\]\[+y{\it arccosh}({\frac
{y}{zu+vw}})-\sqrt {-{\frac {-y+zu+vw}{zu+vw}}}\sqrt {{\frac
{y+zu+vw}{zu+vw}}}zu.
\]

\section{Six-dimensional generalization of the second heavenly
equation}

The second heavenly equation in dimension Dim=6 can be obtained as
generalization of the metric (\ref{dryuma:eq12}).

   It looks as
\begin{equation}\label{dryuma:eq21}{{\it ds}}^{2}=A(\vec x){d{{x}}}^{2}+2\,B(\vec x)d{{x}}d{
{y}}+C(\vec x){d{{y}}}^{2}+2\,E(\vec x)d{{x}}d{{z}}+\]\[+F(\vec
x){d{{z}}}^{2}+2\,H(\vec x)d{{z}}d{{y}}+d{{x}}d{{u}}+
d{{y}}d{{v}}+d{{z}}d{{w}}.
\end{equation}

   The Ricci tensor of such metrics has a fifteen components among which
   are the nine components expressed trough the values
   \[
  P= {\frac {\partial }{\partial u}}E(\vec x)+{\frac {\partial }{
\partial v}}H(\vec x)+{\frac {\partial }{\partial w}}F(\vec x),\]\[Q={\frac {\partial }{\partial u}}B(\vec x)+{\frac
{\partial }{
\partial v}}C(\vec x)+{\frac {\partial }{\partial w}}H(\vec x),\]\[S={\frac {\partial }{\partial u}}A(\vec x)+{\frac
{\partial }{
\partial v}}B(\vec x)+{\frac {\partial }{\partial w}}E(\vec x),
w)
\]
where $\vec x=(x,y,z,u,v,w)$.

     Equating these expressions to zero we get the equations for determination of the components
     of the metrics (\ref{dryuma:eq21}).
\begin{equation}\label{dryuma:eq22}
   {\frac {\partial }{\partial u}}E(x,y,z,u,v,w)+{\frac {\partial }{
\partial v}}H(x,y,z,u,v,w)+{\frac {\partial }{\partial w}}F(x,y,z,u,v,
w)=0,\]\[{\frac {\partial }{\partial u}}B(x,y,z,u,v,w)+{\frac
{\partial }{
\partial v}}C(x,y,z,u,v,w)+{\frac {\partial }{\partial w}}H(x,y,z,u,v,
w)=0,\]\[{\frac {\partial }{\partial u}}A(x,y,z,u,v,w)+{\frac
{\partial }{
\partial v}}B(x,y,z,u,v,w)+{\frac {\partial }{\partial w}}E(x,y,z,u,v.
w)=0. \end{equation}

   This system of equation has a solutions  depended from the arbitrary functions.

      As example the solution of the system depended from one arbitrary
       function lead to the six-dimensional space with the metrics
\begin{equation}\label{dryuma:eq23}^{6}ds^2=\left ({\frac {\partial ^{2}}{\partial {w}^{2}}}K(\vec x)
{\frac {\partial ^{2}}{\partial {v}^{2}}}K(\vec x)-\left ({\frac
{\partial ^{2}}{\partial v\partial w}}K(\vec x)\right )^{ 2}\right
){d{{x}}}^{2}\!+\!\]\[+2\,\left ({\frac {\partial ^{2}}{
\partial u\partial w}}K(\vec x){\frac {\partial ^{2}}{
\partial v\partial w}}K(\vec x)\!-\!{\frac {\partial ^{2}}{
\partial {w}^{2}}}K(\vec x){\frac {\partial ^{2}}{
\partial u\partial v}}K(\vec x)\right )d{{x}}d{{y}}\!+\!\]\[\!+\!\left (
{\frac {\partial ^{2}}{\partial {u}^{2}}}K(\vec x) {\frac
{\partial ^{2}}{\partial {w}^{2}}}K(\vec x)\!-\!\left ({\frac
{\partial ^{2}}{\partial u\partial w}}K(\vec x)\right )^{2} \right
){d{{y}}}^{2}\!+\!\]\[+2\,\left ({\frac {\partial ^{2}}{
\partial v\partial w}}K(\vec x){\frac {\partial ^{2}}{
\partial u\partial v}}K(\vec x)\!-\!{\frac {\partial ^{2}}{
\partial u\partial w}}K(\vec x){\frac {\partial ^{2}}{
\partial {v}^{2}}}K(\vec x)\right )d{{x}}d{{z}}\!+\!\]\[\!+\!\left ({
\frac {\partial ^{2}}{\partial {v}^{2}}}K(\vec x){\frac {
\partial ^{2}}{\partial {u}^{2}}}K(\vec x)\!-\!\left ({\frac {
\partial ^{2}}{\partial u\partial v}}K(\vec x)\right )^{2}\right
){d{{z}}}^{2}\!+\!\]\[\!+\!2\,\left ({\frac {\partial
^{2}}{\partial u
\partial w}}K(\vec x){\frac {\partial ^{2}}{\partial u
\partial v}}K(\vec x)\!-\!{\frac {\partial ^{2}}{\partial {u}^
{2}}}K(\vec x){\frac {\partial ^{2}}{\partial v\partial w }}K(\vec
x)\right )d{{z}}d{{y}}\!+\!\]\[\!+\!d{{x}}d{{u}}+d{{y}}d{{v}}+d
{{z}}d{{w}}
\end{equation}
with arbitrary function $K(\vec x)=K(x,y,z,u,v,w)$.

   This metric is a Ricci-flat in a simplest case
   \[
   K(x,y,z,u,v,w)=F(x,y,z,u)w+v
\]
   where the function $F(x,y,z,u)$ is solution of the equation
\[
\left ({\frac {\partial ^{2}}{\partial u\partial
x}}F(x,y,z,u)\right ) {\frac {\partial ^{2}}{\partial
{u}^{2}}}F(x,y,z,u)+\left ({\frac {
\partial }{\partial u}}F(x,y,z,u)\right ){\frac {\partial ^{3}}{
\partial {u}^{2}\partial x}}F(x,y,z,u)=0.
\]

    In more general case the solution of the system (\ref{dryuma:eq22}) lead to the space with the metric
\begin{equation}\label{dryuma:eq24}
{{\it ^{6}ds}}^{2}=\left ({\frac {\partial ^{2}}{\partial
v\partial w}}N(\vec x)\right ){{\it dx}}^{2}+\left ({\frac
{\partial ^{2} }{\partial {w}^{2}}}M(\vec x)-{\frac {\partial
^{2}}{
\partial u\partial w}}N(\vec x)-{\frac {\partial ^{2}}{
\partial v\partial w}}L(\vec x)\right ){\it dx}\,{\it dy}+\]\[+
\left ({\frac {\partial ^{2}}{\partial {v}^{2}}}L(\vec x)-{\frac
{\partial ^{2}}{\partial u\partial v}}N(\vec x)-{ \frac {\partial
^{2}}{\partial v\partial w}}M(\vec x)\right ){ \it dx}\,{\it
dz}+{\it dx}\,{\it du}+\left ({\frac {\partial ^{2}}{
\partial u\partial w}}L(\vec x)\right ){{\it dy}}^{2}+\]
\[+\left ({\frac {\partial ^{2}}{\partial {u}^{2}}}N(\vec x)-{
\frac {\partial ^{2}}{\partial u\partial v}}L(\vec x)-{ \frac
{\partial ^{2}}{\partial u\partial w}}M(\vec x)\right ){ \it
dy}\,{\it dz}+{\it dy}\,{\it dv}+\left ({\frac {\partial ^{2}}{
\partial u\partial v}}M(\vec x)\right ){{\it dz}}^{2}+{\it dz}\,{
\it dw}, \end{equation}
 where  $\vec x=(x,y,z,u,v,w)$ and $L(\vec x),~M(\vec x),~N(\vec x)$ are an arbitrary functions.

    The metric (\ref{dryuma:eq24})  has a six nonzero components of the Ricci tensor.

     The condition on the space (\ref{dryuma:eq24}) to be  a Ricci-flat lead to the overdetermined
     system of equations for the functions $L(\vec x),~M(\vec x),~N(\vec
     x)$
\[
     R_{xx}=0,\quad R_{xy}=0,\quad R_{xz}=0,\quad R_{yy}=0,\quad R_{yz}=0,\quad R_{zz}=0,
     \]
which is the six dimensional generalization of the
four-dimensional Plebanski the second
    heavenly equation.

     In the case
     \[L(x,y,z,u,v,w)={\it a2}(x,y,z){u}^{2}v+{\it a3}(x,y,z)u{v}^{2}+{\it a5
}(x,y,z){u}^{2}w+\]\[+{\it a6}(x,y,z)u{w}^{2}+{\it
a8}(x,y,z){w}^{2}v+{\it a9}(x,y,z)w{v}^{2} ,
\]
\[
M(x,y,z,u,v,w)={\it b2}(x,y,z){u}^{2}v+{\it
b3}(x,y,z)u{v}^{2}+{\it b5 }(x,y,z){u}^{2}w+\]\[+{\it
b6}(x,y,z)u{w}^{2}+{\it b8}(x,y,z){w}^{2}v+{\it
b9}(x,y,z)w{v}^{2},
\]
\[N(x,y,z,u,v,w)={\it c2}(x,y,z){u}^{2}v+{\it c3}(x,y,z)u{v}^{2}+{\it c5
}(x,y,z){u}^{2}w+\]\[+{\it c6}(x,y,z)u{w}^{2}+{\it
c8}(x,y,z){w}^{2}v+{\it c9}(x,y,z)w{v}^{2} ,
\]
where $a_i,~b_i,~c_i$ are an arbitrary functions we get the space
which is the  Riemann extension of corresponding affinely
connected three-dimensional  space
(\cite{dryuma:dr2},\cite{dryuma:dr4})
\[
ds^2=-2\Gamma^k_{ij}dx^idx^j\Psi_k+2dx^l\Psi_l.
\]

One part of geodesic of such type of the space is defined by
\[{\frac {
d^{2}}{d{s}^{2}}}x(s)-2\,{\it b2}(x,y,z)\left ({\frac
{d}{ds}}z(s)\right )^{2}+\]\[+\left ( \left (2\,{\it
a2}(x,y,z)+2\,{\it b5}(x,y,z)\right ){\frac {d}{ds}}y(s )+\left
(-2\,{\it a3}(x,y,z)+2\,{\it c2}(x,y,z)\right ){\frac {d}{ds}}
x(s)\right ){\frac {d}{ds}}z(s)-\]\[-2\,{\it a5}(x,y,z)\left
({\frac {d}{ds }}y(s)\right )^{2}+\left (-2\,{\it
b6}(x,y,z)+2\,{\it c5}(x,y,z) \right )\left ({\frac
{d}{ds}}x(s)\right ){\frac {d}{ds}}y(s) =0,
\]
\[
{\frac {d^{2}} {d{s}^{2}}}y(s)-2\,{\it b3}(x,y,z)\left ({\frac
{d}{ds}}z(s)\right )^{2}+\]\[+\left ( \left (-2\,{\it
c2}(x,y,z)+2\,{\it a3}(x,y,z)\right ){\frac {d}{ds}}y( s)+\left
(2\,{\it c3}(x,y,z)+2\,{\it b9}(x,y,z)\right ){\frac {d}{ds}}
x(s)\right ){\frac {d}{ds}}z(s)+\]\[+\left (-2\,{\it
b8}(x,y,z)+2\,{\it a9} (x,y,z)\right )\left ({\frac
{d}{ds}}x(s)\right ){\frac {d}{ds}}y(s)-2 \,{\it c9}(x,y,z)\left
({\frac {d}{ds}}x(s)\right )^{2}=0,
\]
\\[0.5mm]\[
\left (\left (-2\,{\it c5}(x,y,z)+2\,{\it b6}(x,y,z)\right ){\frac
{d} {ds}}y(s)+\left (2\,{\it b8}(x,y,z)-2\,{\it a9}(x,y,z)\right
){\frac { d}{ds}}x(s)\right ){\frac {d}{ds}}z(s)-\]\[-2\,{\it
a6}(x,y,z)\left ({ \frac {d}{ds}}y(s)\right )^{2}+\left (2\,{\it
c6}(x,y,z)+2\,{\it a8}(x ,y,z)\right )\left ({\frac
{d}{ds}}x(s)\right ){\frac {d}{ds}}y(s)-2\, {\it c8}(x,y,z)\left
({\frac {d}{ds}}x(s)\right )^{2}+\]\[+{\frac {d^{2}}{d
{s}^{2}}}z(s)
 =0.
\]

    Another part of geodesic is defined by the linear system of equations
    \[
    \frac{d^2\vec \Psi}{ds^2}+A(x,y,z)\frac{d\vec\Psi}{ds}+B(x,y,z)\vec \Psi=0
    \]
    with some $3\times3$ matrix-functions $A,~B$ and $\vec\Psi=(u,v,w)$.

    In considered case the conditions $R_{ij}=0$ lead to the system of equations for the
    coefficients  $a_i,~b_i,~c_i$
\[R_{xx}=-4\,{\it a8}(x,y,z){\it c3}(x,y,z)-4\,{
\frac {\partial }{\partial z}}{\it c8}(x,y,z)-8\,{\it
c8}(x,y,z){\it a3}(x,y,z)-\]\[-4\,{\it c6}(x,y,z){\it
b9}(x,y,z)-4\,{\it a8}(x,y,z){\it b9 }(x,y,z)+8\,{\it
b8}(x,y,z){\it a9}(x,y,z)+8\,{\it c9}(x,y,z){\it c5}(
x,y,z)+\]\[+8\,{\it c8}(x,y,z){\it c2}(x,y,z)-4\,{\it
c6}(x,y,z){\it c3}(x, y,z)-8\,{\it c9}(x,y,z){\it
b6}(x,y,z)-\]\[-4\,\left ({\it a9}(x,y,z) \right )^{2}-4\,\left
({\it b8}(x,y,z)\right )^{2}-4\,{\frac {
\partial }{\partial y}}{\it c9}(x,y,z)=0,
\]
 \[
R_{yy}= 8\,{\it a5}(x,y,z){\it a9}(x,y,z)-8\,{\it a6}(x,y,z){\it
c2}(x,y,z)-8 \,{\it a5}(x,y,z){\it b8}(x,y,z)-\]\[-4\,{\it
a8}(x,y,z){\it b5}(x,y,z)-4\, {\it a8}(x,y,z){\it
a2}(x,y,z)+8\,{\it a6}(x,y,z){\it a3}(x,y,z)-\]\[-4\,{ \frac
{\partial }{\partial z}}{\it a6}(x,y,z)-4\,{\frac {\partial }{
\partial x}}{\it a5}(x,y,z)-4\,\left ({\it b6}(x,y,z)\right )^{2}-4\,
\left ({\it c5}(x,y,z)\right )^{2}-\]\[-4\,{\it c6}(x,y,z){\it
a2}(x,y,z)+8 \,{\it b6}(x,y,z){\it c5}(x,y,z)-4\,{\it
c6}(x,y,z){\it b5}(x,y,z)=0,
\]
\[
R_{zz}=8\,{\it b2}(x,y,z){\it b8}(x,y,z)-4\,{\it c3}(x,y,z){\it
b5}(x,y,z)-8 \,{\it b2}(x,y,z){\it a9}(x,y,z)-\]\[-8\,{\it
b3}(x,y,z){\it c5}(x,y,z)-4\, {\it b9}(x,y,z){\it
b5}(x,y,z)+8\,{\it a3}(x,y,z){\it c2}(x,y,z)-\]\[-4\,{ \it
b9}(x,y,z){\it a2}(x,y,z)+8\,{\it b3}(x,y,z){\it b6}(x,y,z)-4\,{
\it c3}(x,y,z){\it a2}(x,y,z)-\]\[-4\,\left ({\it a3}(x,y,z)\right
)^{2}-4 \,{\frac {\partial }{\partial x}}{\it b2}(x,y,z)-4\,{\frac
{\partial } {\partial y}}{\it b3}(x,y,z)-4\,\left ({\it
c2}(x,y,z)\right )^{2}=0,
\]
\[
R_{xy}=4\,{\it c8}(x,y,z){\it a2}(x,y,z)+4\,{\it b6}(x,y,z){\it
a9}(x,y,z)-4 \,{\it b6}(x,y,z){\it b8}(x,y,z)+\]\[+4\,{\it
c5}(x,y,z){\it b8}(x,y,z)+2\, {\frac {\partial }{\partial y}}{\it
a9}(x,y,z)+2\,{\frac {\partial }{
\partial z}}{\it a8}(x,y,z)+2\,{\frac {\partial }{\partial z}}{\it c6}
(x,y,z)+\]\[+4\,{\it c8}(x,y,z){\it b5}(x,y,z)+2\,{\frac {\partial
}{
\partial x}}{\it c5}(x,y,z)+4\,{\it a6}(x,y,z){\it b9}(x,y,z)-8\,{\it
c9}(x,y,z){\it a5}(x,y,z)-\]\[-\,{\it c5}(x,y,z){\it
a9}(x,y,z)-2\,{\frac {\partial }{\partial x}}{\it
b6}(x,y,z)+4\,{\it a6}(x,y,z){\it c3}(x,y ,z)-2\,{\frac {\partial
}{\partial y}}{\it b8}(x,y,z)=0,
\]
\[
R_{xz}=4\,{\it a3}(x,y,z){\it b8}(x,y,z)+2\,{\frac {\partial
}{\partial x}}{ \it c2}(x,y,z)+4\,{\it b3}(x,y,z){\it
a8}(x,y,z)-\]\[-8\,{\it c8}(x,y,z){ \it b2}(x,y,z)+4\,{\it
c2}(x,y,z){\it a9}(x,y,z)+4\,{\it c9}(x,y,z){ \it
a2}(x,y,z)-4\,{\it c2}(x,y,z){\it b8}(x,y,z)+\]\[+4\,{\it
b3}(x,y,z){ \it c6}(x,y,z)-4\,{\it a3}(x,y,z){\it
a9}(x,y,z)+2\,{\frac {\partial } {\partial y}}{\it
c3}(x,y,z)+2\,{\frac {\partial }{\partial z}}{\it b8
}(x,y,z)+\]\[+2\,{\frac {\partial }{\partial y}}{\it
b9}(x,y,z)-2\,{\frac {
\partial }{\partial x}}{\it a3}(x,y,z)-2\,{\frac {\partial }{\partial
z}}{\it a9}(x,y,z)+4\,{\it c9}(x,y,z){\it b5}(x,y,z)=0,
\]
\[
R_{yz}=4\,{\it a5}(x,y,z){\it b9}(x,y,z)+4\,{\it b2}(x,y,z){\it
c6}(x,y,z)+4 \,{\it a5}(x,y,z){\it c3}(x,y,z)-\]\[-4\,{\it
a3}(x,y,z){\it b6}(x,y,z)+4\, {\it b2}(x,y,z){\it
a8}(x,y,z)-4\,{\it c2}(x,y,z){\it c5}(x,y,z)-\]\[-8\,{ \it
a6}(x,y,z){\it b3}(x,y,z)+4\,{\it a3}(x,y,z){\it c5}(x,y,z)-2\,{
\frac {\partial }{\partial z}}{\it c5}(x,y,z)+\]\[+2\,{\frac
{\partial }{
\partial y}}{\it a3}(x,y,z)+2\,{\frac {\partial }{\partial x}}{\it a2}
(x,y,z)-2\,{\frac {\partial }{\partial y}}{\it
c2}(x,y,z)+2\,{\frac {
\partial }{\partial x}}{\it b5}(x,y,z)+\]\[+2\,{\frac {\partial }{\partial
z}}{\it b6}(x,y,z)+4\,{\it c2}(x,y,z){\it b6}(x,y,z) =0,
\]
 which can be of used for the extension of  the Liouville
    theory of invariants of the second order ODE's  on a three dimensional case.

    In particular the description of the projectively flat
    structures defined by the solutions of  nonlinear p.d.e.'s connected with
    corresponding systems of ODE's is possible.

\section{Eight-dimensional generalization}

     By analogy  the eight-dimensional generalization of the second
     Heavenly equation can be obtained.

     The metrics of corresponding space has the form

\begin{equation}\label{dryuma:eq25}
^{8}ds^2=\left (-{\frac {\partial ^{2}}{\partial
{q}^{2}}}\omega(\vec x)-{\frac {\partial ^{2}}{\partial
{u}^{2}}}\alpha(\vec x)-{ \frac {\partial ^{2}}{\partial
{v}^{2}}}\delta(\vec x)\right ){d{{x}}}^{2}+2\,\left ({\frac
{\partial ^{2}}{\partial p\partial q}} \omega(\vec x)\right
)d{{x}}d{{y}}+\]\[+2\,\left ({\frac {
\partial ^{2}}{\partial p\partial u}}\alpha(\vec x)\right )d
{{x}}d{{z}}+2\,\left ({\frac {\partial ^{2}}{\partial p\partial
v}} \delta(\vec x)\right )d{{x}}d{{t}}+\left (-{\frac {
\partial ^{2}}{\partial {p}^{2}}}\omega(\vec x)-{\frac {
\partial ^{2}}{\partial {u}^{2}}}\beta(\vec x)-{\frac {
\partial ^{2}}{\partial {v}^{2}}}\mu(\vec x)\right ){d{{y}}}
^{2}+\]\[+2\,\left ({\frac {\partial ^{2}}{\partial q\partial
u}}\beta(\vec x)\right )d{{y}}d{{z}}+2\,\left ({\frac {\partial
^{2}}{
\partial q\partial v}}\mu(\vec x)\right )d{{y}}d{{t}}+
\left (-{\frac {\partial ^{2}}{\partial {v}^{2}}}\nu(\vec x)-
{\frac {\partial ^{2}}{\partial {q}^{2}}}\beta(\vec x)-{ \frac
{\partial ^{2}}{\partial {p}^{2}}}\alpha(\vec x)\right
){d{{z}}}^{2}+\]\[+2\,\left ({\frac {\partial ^{2}}{\partial
u\partial v}} \nu(\vec x)\right )d{{z}}d{{t}}+\left (-{\frac
{\partial ^{ 2}}{\partial {p}^{2}}}\delta(\vec x)-{\frac {\partial
^{2}}{
\partial {q}^{2}}}\mu(\vec x)-{\frac {\partial ^{2}}{
\partial {u}^{2}}}\nu(\vec x)\right ){d{{t}}}^{2}+\]\[+d{{x}}d{
{p}}+d{{q}}d{{y}}+d{{z}}d{{u}}+d{{t}}d{{v}},
 \end{equation}  depending on the six arbitrary functions and $\vec x=(x,y,z,t,p,q,u,v)$.

     Remark that a Ricci tensor of the metrics (\ref{dryuma:eq25}) has a
     ten  nonzero components.

        The generalization of the second Heavenly equations corresponds the
        system from a six equations on the coefficients of the metrics (\ref{dryuma:eq25}).

\section{Acknowledgement}

     The research was partially supported by the Grant 06.01 CRF of HCSTD ASM and the RFBR Grant.

\end{document}